\documentclass[letterpaper, 10 pt, conference]{ieeeconf}  

\IEEEoverridecommandlockouts                              
\usepackage{amsmath}
\usepackage{amsthm}
\usepackage{amsfonts}
\usepackage{xcolor}
\usepackage{subfig}
\usepackage{graphicx}

\usepackage{hyperref}   
\newtheorem{assumption}{Assumption}
\newtheorem{problem}{Problem}
\newtheorem{lemmma}{Lemma}
\newcommand{\E}{\operatorname{\mathbb{E}}} 

\newcommand{\beq}[1]{\begin{align*}\label{eq:#1}}
\newcommand{\eeq}{\end{align*}}

\makeatletter
\newcommand{\xMapsto}[2][]{\ext@arrow 0599{\Mapstofill@}{#1}{#2}}
\def\Mapstofill@{\arrowfill@{\Mapstochar\Relbar}\Relbar\Rightarrow}
\makeatother

\newtheorem{example}{Example}
\newtheorem{theorem}{Theorem}
\newtheorem{remark}{Remark}
\title{\LARGE \bf
Beyond Quadratic Costs in LQR: Bregman Divergence Control
}

\author{Babak Hassibi, Joudi Hajar, and Reza Ghane
\thanks{The authors are in the Department of Electrical Engineering, Caltech. Emails:
        {\tt\small \{hassibi,jhajar,rghanekh\}@caltech.edu}}}
\usepackage{comment}
\begin{document}

\maketitle
\thispagestyle{empty}
\pagestyle{empty}

\begin{abstract}

In the past couple of decades, the use of ``non-quadratic" convex cost functions has revolutionized signal processing, machine learning, and statistics, allowing one to customize solutions to have desired structures and properties. However, the situation is not the same in control where the use of quadratic costs still dominates, ostensibly because determining the ``value function", i.e., the optimal expected cost-to-go, which is critical to the construction of the optimal controller, becomes computationally intractable as soon as one considers general convex costs. As a result, practitioners often resort to heuristics and approximations, such as model predictive control that only looks a few steps into the future. In the quadratic case, the value function is easily determined by solving Riccati equations. In this work, we consider a special class of convex cost functions constructed from Bregman divergence and show how, with appropriate choices, they can be used to fully extend the framework developed for the quadratic case. The resulting optimal controllers are infinite horizon, come with stability guarantees, and have state-feedback, or estimated state-feedback, laws. They exhibit a much wider range of behavior than their quadratic counterparts since the feedback laws are nonlinear. The approach can be applied to several cases of interest, including safety control, sparse control, and bang-bang control. 

\end{abstract}

\section{Introduction}

Optimal stochastic control theory has traditionally centered around quadratic cost functions, as exemplified by the Linear Quadratic Gaussian (LQG) framework \cite{kalman1960contributions,yakubovich1962solution, popov1964hyperstability}. This methodology optimizes quadratic costs on both system states and control inputs, leveraging the mathematical tractability of quadratic forms to ensure computational efficiency and system stability. However, the exclusive reliance on quadratic costs imposes significant limitations, particularly in applications requiring more sophisticated performance metrics and structural constraints.

In contrast, disciplines such as machine learning, signal processing, and statistics have effectively extended beyond quadratic losses by incorporating a broader class of convex cost functions. Techniques including $l_1$-regularization for sparsity promotion \cite{donoho2005sparse,candes2006robust}, nuclear norm minimization for low-rank matrix recovery \cite{recht2008necessary,recht2010guaranteed}, and cross-entropy loss for classification \cite{shore1981properties,shore1982minimum, ruby2020binary} tasks illustrate the enhanced flexibility and performance achievable with non-quadratic convex costs. These advancements enable the imposition of complex structural constraints and the recovery of intricate models that quadratic costs alone cannot accommodate.

\vspace{0.5em} Despite the demonstrated advantages in these fields, the integration of non-quadratic convex costs into stochastic control remains limited. The challenges hindering this integration include increased computational complexity and, more importantly, the difficulty in managing stochastic disturbances in the non-quadratic setting. While quadratic control problems benefit from efficient dynamic programming solutions with linear scaling relative to the planning horizon, non-quadratic convex control problems typically exhibit a complexity that grows polynomially with the time horizon, rendering them computationally intractable for extended (or infinite) horizons. Additionally, incorporating noise into the system complicates the computation of the optomal expected cost-to-go, as non-quadratic cost functions often lack closed-form solutions for the value function, this necessitating inefficient Monte Carlo methods that can only be run for a few time steps in the future (receding horizon control). 

\vspace{0.5em} 
In our companion paper~\cite{ourpaper3}, we addressed some of these challenges by extending classical robust \(H_\infty\) control from quadratic to strictly convex cost functions through a Bregman–divergence approach. Concurrently, the present work develops a complementary theory for \(H_2\) control by introducing a novel framework that extends classical quadratic \(H_2\) (stochastic) control techniques to accommodate a wider range of convex cost functions through the utilization of Bregman divergence. Bregman divergence serves as a natural generalization of quadratic forms, preserving key properties, such as ``law of cosines" and ``completion-of-squares", while allowing for greater flexibility in defining cost landscapes. Most critically, by reformulating the control problem in terms of Bregman divergence, we can effectively separate the deterministic and stochastic components of the cost, thereby facilitating the computation of expected costs and enabling the design of optimal controllers even in the presence of noise.

\vspace{0.5em} The primary contributions of this work are as follows:

\textbf{Bregman Divergence Control Framework:} We develop a comprehensive framework that leverages Bregman divergence to extend optimal stochastic control beyond quadratic costs, providing a systematic approach to incorporate diverse non-quadratic convex cost penalties on both the state and control. 
To circumvent the complexity of nested expectations, we insist that the value function (or optimal expected cost-to-go) be quadratic in the state. While this allows for the expectation to be computed in closed form, it allows one to only freely choose one of the state or control penalty functions--the other one will follow.

\textbf{Generalized Riccati-Like Equation:} We derive a counterpart to the Riccati equation tailored to Bregman divergence. This generalized equation establishes conditions under which the convex functions defining the state and control costs, along with a positive definite matrix, ensure the existence of optimal nonlinear state feedback controllers.

\textbf{Stability:} Our framework, the sum of the state cost and positive definite value function, yields a non-quadratic Lyapunov function that ensures stability. 

\textbf{Illustrative Examples:} We mention ``safety control" and ``bang-bang" control as examples demonstrating the versatility and effectiveness of Bregman divergence control. However, in the interest of space, we only study the case of ``elastic net" control, which puts a mixed $\ell_1/\ell_2$ cost on the state. We show that the resulting controller sparsifies the state (pushing it to zero faster) and exhibits behavior qualitatively different from standard LQG controllers. 

\textbf{Computational Feasibility:} By reformulating the control problem with Bregman divergence, we mitigate the computational challenges associated with non-quadratic costs, enabling scalable solutions for infinite horizon problems and complex stochastic environments.

Through these contributions, this paper bridges a critical gap in control theory, enabling the incorporation of diverse convex cost functions and significantly broadening the applicability of stochastic control. The ensuing sections detail the theoretical foundations, algorithmic implementations, and practical applications of linear Bregman divergence control, highlighting its potential to advance both the theoretical and practical aspects of modern control systems.


\section{Preliminaries}
\subsection{Notations}
The set of real numbers is denoted by \(\mathbb{R}\), and \(\mathbb{E}\) represents the expectation operator. The \(\ell_2\)-norm is written as \(\|\cdot\|_2\). For any matrix \(X\), its transpose is denoted by \(X^\top\) and its Moore–Penrose pseudoinverse by \(X^\dagger\). For symmetric matrices \(A\) and \(B\), we write \(A \succeq B\) if \(A - B\) is positive semidefinite, and \(A \succ B\) if \(A - B\) is positive definite.

For a function \(f:\mathbb{R}^n \to \mathbb{R}\), its Fenchel dual is defined as 
\[
f^\ast(y) = \sup_{x\in\mathbb{R}^n} \{ x^\top y - f(x) \}.
\]
Moreover, \(\nabla f(\cdot)\) denotes the gradient of \(f\), \(\nabla^2 f(\cdot)\) its Hessian, and \(\nabla^{-1} f(\cdot)\) the inverse of the gradient operator.

\subsection{Linear Quadratic Regulator (LQR) for LTI Systems}
Consider the discrete-time LTI system:
\begin{align}
    x_{k+1} = A x_k + B u_k + w_k,
    \label{eq:system_dynamics_combined}
\end{align}
where \( x_k \in \mathbb{R}^n \) is the state, \( u_k \in \mathbb{R}^m \) is the control input, \( A \in \mathbb{R}^{n \times n} \) and \( B \in \mathbb{R}^{n \times m} \) are constant system matrices, and \( w_k \) is a zero-mean noise with covariance \( W \). The objective is to design a controller that minimizes a quadratic cost.

\vspace{0.5em}
\noindent \textbf{Finite-Horizon LQR:} For a finite planning horizon \( N \), the goal is to determine the sequence of control inputs \( \{u_0, u_1, \dots, u_{N-1}\} \), each of which is a {\em strictly causal} function of the noise inputs, that minimizes the expected cost:
\begin{align}
    \min_{\{u_k\}} \mathbb{E}_{\{w_0,\dots, w_{N-1}\}} \Biggl[ \!x_N^\top P_N x_N \!+ \!\sum_{k=0}^{N-1} \!\Bigl(\! x_k^\top Q_k x_k + u_k^\top R_k u_k \Bigr) \!\Biggr]
    \label{eq:quadratic_cost_finite}
\end{align}
where \( P_N,Q_k \in \mathbb{R}^{n \times n} \) (with \( P_N,Q_k \succeq 0 \)) and \( R_k \in \mathbb{R}^{m \times m} \) (with \( R_k \succeq 0 \)) for all $k\in(0\cdots N-1)$ are the state and control weighting matrices, respectively. 

The optimal control policy, derived using dynamic programming \cite{bertsekas1995}, is the linear state-feedback:
    $u_k^\ast(x_k) = -\left(B^\top K_{k+1} B + R_k\right)^{-1} B^\top K_{k+1} A\, x_k,$
    \label{eq:lqr_control_finite}
with \( K_k \in \mathbb{R}^{n \times n} \) satisfying the Riccati recursion:
$K_k = Q_k + A^\top \Bigl( K_{k+1} - K_{k+1} B \left( B^\top K_{k+1} B + R_k \right)^{-1} B^\top K_{k+1} \Bigr) A,$
and with the terminal condition \( K_N = P_N \).

\vspace{0.5em}
\noindent \textbf{Infinite-Horizon LQR:} As the planning horizon \( N \) tends to infinity, the finite-horizon problem converges (under stabilizability and detectability conditions) to the infinite-horizon formulation, where the objective is to minimize:
\begin{align}
    \min_{\{u_k\}} \lim_{N\rightarrow\infty} \mathbb{E} \left[ \frac{1}{N}\sum_{k=0}^{N-1} \left( x_k^\top Q x_k + u_k^\top R u_k \right) \right].
    \label{eq:quadratic_cost_infinite}
\end{align}
In this setting, the optimal control law becomes time-invariant:
$u^\ast(x) = -\left(B^\top K B + R\right)^{-1} B^\top K A\, x,$
where \( K \in \mathbb{R}^{n \times n} \) is the unique positive semi-definite solution to the discrete-time algebraic Riccati equation (DARE): 
\begin{align*}
    K = Q + A^\top \Bigl( K - K B \left( B^\top K B + R \right)^{-1} B^\top K \Bigr) A.
\end{align*}
Solving the DARE guarantees that the closed-loop system matrix \( A - B\left(B^\top K B + R\right)^{-1} B^\top K A \) is stable.

\subsection{Bregman Divergence}\label{sec:prelim_breg}

 Given a strictly convex and differentiable function $\phi: \mathbb{R}^n \rightarrow \mathbb{R}$, the Bregman divergence between two points $x, y \in \mathbb{R}^n$ is defined as:
\begin{equation}
    D_\phi(x, y) = \phi(x) - \phi(y) - \nabla \phi(y)^\top (x - y),
    \label{eq:bregman_divergence}
\end{equation}
where $\nabla \phi(y)$ denotes the gradient of $\phi$ evaluated at $y$. This divergence quantifies the difference between the value of the convex function at $x$ and its first-order Taylor approximation around $y$ evaluated at $x$. While $D_\phi(x, y)$ shares some properties with a distance metric, it is neither symmetric nor does it satisfy the triangle inequality---thus it is classified as a divergence rather than a metric.

\begin{example}\label{example:1}
    \emph{Squared Euclidean Distance:} Let $\phi(x) = \frac{1}{2} \|x\|_2^2$. Substituting into the definition of Bregman divergence yields:
$D_\phi(x, y) = \frac{1}{2} \|x\|_2^2 - \frac{1}{2} \|y\|_2^2 - y^\top (x - y) = \frac{1}{2} \|x - y\|_2^2.$
\end{example}
\begin{example}
    \emph{Kullback-Leibler (KL) Divergence:} Consider the negative entropy function $\phi(x) = \sum_{i=1}^n x_i \log x_i - x_i$. The corresponding Bregman divergence is the KL divergence:$D_\phi(x, y) = \sum_{i=1}^n x_i \log \frac{x_i}{y_i} - x_i + y_i,$ which is a common measure for the difference between two probability distributions.
\end{example}

Since Bregman divergence retains the quadratic and higher-order terms in $\phi(\cdot)$, it shares many properties similar to quadratics. 

\begin{lemmma}
    The following holds:
    \begin{itemize}
        \item \emph{\textbf{Non-Negativity}}: $D_\phi(x, y) \geq 0$ for all $x, y \in \mathbb{R}^n$, with equality if and only if $x = y$.
        \item \emph{\textbf{Convexity}}: For any fixed $y$, $D_\phi(\cdot,y)$ is convex.
        \item \emph{\textbf{Law of Cosines:}} For any three points $x, y, z \in \mathbb{R}^n$, the following identity holds:
        \begin{align}
            D_\phi(x, y) = &D_\phi(x, z)+ D_\phi(z, y) \nonumber \\ & - \left( \nabla \phi(y) - \nabla \phi(z) \right)^\top (x - z).
            \label{eq:law_of_cosines}
        \end{align}
       
        \item  \emph{\textbf{Completion of Squares:}} For two strictly convex functions $\phi_1$ and $\phi_2$, and points $x, y, z \in \mathbb{R}^n$, the following identity holds:
        \begin{align}
            D_{\phi_1}(x, y) + D_{\phi_2}(x, z) = &D_{\phi_1 + \phi_2}\left(x, x^\ast\right)  + D_{\phi_1}\left(x^\ast, y\right) \nonumber \\ & + D_{\phi_2}\left(x^\ast, z\right),
            \label{eq:completion_of_squares}
        \end{align}
        where $x^\ast$ satisfies:
        \begin{equation}
            \nabla (\phi_1 + \phi_2)(x^\ast) = \nabla \phi_1(y) + \nabla \phi_2(z).
        \end{equation}
    \end{itemize}
\end{lemmma}

These properties make Bregman divergence a versatile tool in various optimization applications.

\vspace{2mm}

An essential property of Bregman divergence in the context of stochastic control is its behavior under expectation. Specifically, the expected divergence between a deterministic point \(x\) and a random variable \(y\) can be decomposed via the following lemma:
\begin{lemmma}\label{lem:expectation_property} (\emph{\textbf{Expectation Property)}}
    Let $x$ and $y$ be deterministic and random variables in $\mathbb{R}^n$, respectively. Then, the expected Bregman divergence is given by:
    \begin{align}
        \mathbb{E} \left[ D_\phi(x, y) \right] = & D_\phi\left(x, \nabla^{-1} \phi\left( \mathbb{E} \left[ \nabla \phi(y) \right] \right) \right)\nonumber \\ & + \mathbb{E} \left[ D_\phi\left( \nabla^{-1} \phi\left( \mathbb{E} \left[ \nabla \phi(y) \right] \right), y \right) \right].
        \label{eq:property_exp}
    \end{align}
\end{lemmma}

Notice that on the rhs of \eqref{eq:property_exp}, the second term is independent of the deterministic variable $x$, and the expectations in the first and second term involve only the random variable $y$.
When \(\phi\) is even and the random variable $y$ has a symmetric distribution, the expression in \ref{eq:property_exp} simplifies further as will be shown later. This tractable handling of expectations is the crux of our approach to designing non-quadratic, yet tractable, stochastic costs in control systems.

\section{The Bregman Divergence Framework and the Bregman Controller}

We aim to move beyond conventional quadratic cost functions (see \eqref{eq:quadratic_cost_finite} and \eqref{eq:quadratic_cost_infinite}) by incorporating more general \emph{convex} penalties on both the state \(x_t\) and the control \(u_t\). A direct substitution of quadratic terms with general convex functions \(q(x_t)\) and \(r(u_t)\) would lead to nested expectations that are not tractable. To overcome this challenge, we replace the original cost with one based on Bregman divergences. The key advantage of this approach is that it enables us to \emph{separate the deterministic} and \emph{stochastic} components in the state update, allowing the expected cost to be computed in closed form and facilitating tractable dynamic programming.

\subsection{Problem Statement}

\begin{problem}\label{pb:bregman}
Let \( r(\cdot) \), \( q(\cdot) \) and $r_k(\cdot),q_k(\cdot)\quad \text{for }k:0\cdots N-1$, $p_N(\cdot)$ be convex, even, and nonnegative functions. Our goal is to determine the optimal sequence of state-feedback control laws \( \{u_k\} \), that minimizes the following Bregman divergence cost:

\vspace{2mm}
\textbf{In the Infinite-Horizon:}
   \begin{align}\label{eq:breg_cost-Inf}
     \min_{\{u_k\}} \lim_{N\rightarrow \infty}\mathbb{E}_{\{w_k\}} \frac{1}{N}\biggl[ \sum_{k=1}^{N-1}  &\Bigr[D_q\Bigl(Ax_{k-1} + Bu_{k-1}, -w_{k-1}\Bigr) \nonumber \\ &+  D_r(u_k, 0)\Bigl]+D_r(u_0, 0)\biggr]     
\end{align}

Here, \(D_r\), and \(D_q\) denote the Bregman divergences associated with the functions \(r\), \(q\), and \(p\), respectively, where:
\begin{itemize}
    \item \(D_r(u_k,0)\) is the cost on the control input $u_k$,
    \item \(D_q(Ax_{k-1}+Bu_{k-1},-w_{k-1})\) penalizes the state $x_k$ at each time step.
\end{itemize}
\textbf{In the Finite-Horizon:}
\begin{align}\label{eq:breg_cost}
\min_{\{u_k\}}  &\E_{\{w_k\}} \nonumber \\&\Biggl[ \sum_{k=1}^{N-1} \! \Big[D_{q_k}\Bigl(\!Ax_{k-1}\!+\!Bu_{k-1},-w_{k-1}\!\Bigr) + D_{r_k}(u_k,0)\Bigr]\nonumber\\
& +\!\!D_{r_0}(u_0,0) + \; D_{p_N}\Bigl(Ax_{N-1}+Bu_{N-1},-w_{N-1}\Bigr)\! \Biggr]
\end{align}
Here, \(D_{r_k}\), and \(D_{q_k}\) are the finite-horizon counterparts of \(D_{r}\) and \(D_{q}\), and $D_{p_N}$ is the Bregman divergence associated with $p_N(\cdot)$ which penalizes the terminal state $x_N$.
\end{problem}
By separating the deterministic update \(Ax_{k-1}+Bu_{k-1}\) from the disturbance \(w_{k-1}\) in $D_q$,  the expectation of this Bregman divergence factorizes nicely using Lemma \ref{lem:expectation_property}. 

\begin{remark}
Choosing \(q(\cdot)\) and \(r(\cdot)\) as quadratic functions recovers the classical LQR cost in \eqref{eq:quadratic_cost_infinite} (cf. Example~\ref{example:1}). 
\end{remark}

\subsection{Assumptions}

\begin{assumption}\label{asum:q}
The functions \(q(\cdot)\) and \(r(\cdot)\) (and their finite horizon counterparts along with $p_N$) are convex, even, and nonnegative. Each of the functions is zero at the origin, with a zero gradient at the origin. (e.g., $q(0)=0,$ $\nabla q(0)=0$.)

\end{assumption}
\begin{assumption}\label{asum:w}
The noise \(w_k\) is drawn from a symmetric distribution about the origin (i.e., zero-mean).
\end{assumption}
\begin{assumption}\label{asum:AB}
The system matrix \(A\) is invertible, and the input matrix \(B\) is full rank.
\end{assumption}

Note that Assumption \ref{asum:q} implies that  the fenchel duals of $q(\cdot)$ and $ r(\cdot)$ inherit similar properties, i.e., $q^\ast(\cdot)$ and $r^\ast(\cdot)$, are convex, even and non negative functions, satisfying the following conditions: $ r^\ast(0) =  q^\ast(0) = 0$ and $\nabla r^\ast(0) = \nabla q^\ast(0) = 0$. Assumption \ref{asum:q}  also allows us to reduce the control cost, \(D_r\bigl(u_t, 0\bigr)\).

\subsection{The Finite Horizon Problem}
We first use the finite horizon problem as a ''motivation'' for the derivation of the infinite horizon problem.
\subsubsection{Dynamic Programming (DP) Recursion for the Bregman Cost}
To solve the optimization problem defined in \eqref{eq:breg_cost}, we employ dynamic programming (DP). Let \(J_k(x_k)\) denote the optimal cost-to-go from state \(x_k\) at time \(k\). The DP recursion is then formulated as follows:
\begin{align}
    J_N(x_N) = D_{p_N}\Bigl(Ax_{N-1}+Bu_{N-1},-w_{N-1}\Bigr), \label{eq:DP_breg_final}
\end{align}
And for $k=1,\!\dots,\!N\!-\!1$:
\begin{align}
    J_k(x_k) = \min_{u_k} \mathbb{E}_{w_k} \biggl[& D_{q_k}\Bigl(Ax_{k-1} +\! Bu_{k-1}, \!-w_{k-1}\Bigr)  \nonumber \\ &+ D_{r_k}(u_k, 0) \biggr] + \!\!J_{k+1}(x_{k+1}),\label{eq:DP_recursive} 
\end{align}
\vspace{-3mm}
\begin{align}   \!\!\!\!\!\!\!\!\!\!\!\!\!\!\!\!\!\!\!\!\!\!\!\!\!\!\!\!\!\!J_0(x_0) = \min_{u_0} \; \Bigl[ D_{r_0}(u_0,0) + J_1(x_1) \Bigr].
\end{align}

Our goal is to show that this DP recursion can be simplified into a deterministic optimization. To illustrate the idea, we start from the DP recursion in \eqref{eq:DP_recursive} and expand $J_{k+1}(x_{k+1})$ at $k=N-1$:
\begin{align}\label{eq:DP_expanded}
    J_{N-1}(x_{N-1}) &= D_{q_{N-1}}\Bigl(Ax_{N-2}+Bu_{N-2},-w_{N-2}\Bigr) \nonumber\\[1mm]
    \quad + \min_{u_{N-1}} & \; \E_{w_{N-1}}\!\Bigl[ D_{p_N}\Bigl(Ax_{N-1}+Bu_{N-1},-w_{N-1}\Bigr) \Bigr] \nonumber \\ & + D_{r_{N-1}}(u_{N-1},0).
\end{align}

\subsubsection{Decomposing the Expected Bregman Divergence}
To handle the expectation in \eqref{eq:DP_expanded}, we use Lemma~\ref{lem:expectation_property} to decompose the expected Bregman divergence between the deterministic term \(Ax_{N-1}+Bu_{N-1}\) and the disturbance \(w_{N-1}\) in a manner analogous to completing the square:
\begin{align}\label{eq:bye}
    &\E_{w_{N-1}}\!\Bigl[ D_{p_{N}}\bigl(Ax_{N-1}+Bu_{N-1},-w_{N-1}\bigr) \Bigr]=
    \nonumber\\ & D_{p_N}\Bigl(Ax_{N-1}+Bu_{N-1},\,\nabla^{-1} p_N\Bigl[\E_{w_{N-1}}\nabla p_N(-w_{N-1})\Bigr]\Bigr) \nonumber\\[1mm]
    & \!+\! \E_{w_{N-1}}\!\Bigl[ D_{p_N}\Bigl(\nabla^{-1} p_N\Bigl[\!\E_{w_{N-1}}\nabla {p_N}(-w_{N-1})\Bigr],-w_{N-1}\Bigr)\! \Bigr].
\end{align}
We denote the second term by
\[
    \resizebox{\columnwidth}{!}{$c_{N-1} :=\E_{w_{N-1}}\!\Bigl[ D_{p_N}\Bigl(\nabla^{-1} {p_N}\Bigl[\E_{w_{N-1}}\nabla {p_N}(-w_{N-1})\Bigr],-w_{N-1}\Bigr) \Bigr],$}
\]
which is independent of the state \(x_{N-1}\) and control \(u_{N-1}\).
\vspace{3mm}

Under Assumption~\ref{asum:w} and given that \(q(\cdot)\) is even, we have
\[
    \E_{w_{N-1}}\bigl[\nabla {p_N}(-w_{N-1})\bigr]=0.
\]
Moreover, Assumption~\ref{asum:q} implies that
\[
    \nabla^{-1} {p_N}\Bigl[\E_{w_{N-1}}\nabla {p_N}(-w_{N-1})\Bigr] = \nabla {p_N}^\ast(0)=0.
\]
Thus, the decomposition in \eqref{eq:bye} simplifies to
\begin{align}\label{eq:bregExpect2}
     &\E_{w_{N-1}}\Bigl[ D_{p_N}\bigl(Ax_{N-1}+Bu_{N-1},-w_{N-1}\bigr) \Bigr]
    \nonumber\\& =D_{p_N}\bigl(Ax_{N-1}+Bu_{N-1},0\bigr) + c_{N-1},
\end{align}
or equivalently,
\begin{align}\label{eq:bregExpect3}
    & \E_{w_{N-1}}\Bigl[ D_{p_N}\bigl(Ax_{N-1}+Bu_{N-1},-w_{N-1}\bigr) \Bigr]
   \nonumber \\ &= {p_N}\bigl(Ax_{N-1}+Bu_{N-1}\bigr) + c_{N-1}
\end{align}
since \({p_N}(0)=0\) and \(\nabla {p_N}(0)=0\) (by Assumption~\ref{asum:q}).
\vspace{3mm}

This result shows that the expected Bregman divergence separates into a term that depends solely on the deterministic dynamics \(Ax_{N-1}+Bu_{N-1}\) and an additive constant \(c_{N-1}\) capturing the influence of the disturbance. This separation is key because it allows us to replace the stochastic optimization by an equivalent deterministic problem.

\subsubsection{Simplified DP Recursion}
Substituting \eqref{eq:bregExpect3} into \eqref{eq:DP_expanded} yields the following optimization for \(J_{N-1}(x_{N-1})\):
\begin{align}
    &J_{N-1}(x_{N-1}) = D_{q_{N-1}}\Bigl(Ax_{N-2}+Bu_{N-2},-w_{N-2}\Bigr) \nonumber \\&+ \min_{u_{N-1}} \Bigl[ r_{N-1}(u_{N-1}) + p_{N}\bigl(Ax_{N-1}+Bu_{N-1}\bigr) \Bigr] + c_{N-1}.\label{eq:DP_simplified}
\end{align}
\vspace{-4mm}

Since the constant \(c_{N-1}\) does not affect the minimization, it can be dropped from the optimization.

Let us define the cost-to-go function at time \(N-1\) by
\begin{align}\label{eq:def_quad_I}
    &\min_{u_{N-1}} \biggl[ r_{N-1}(u_{N-1})\nonumber\\
    &+ p_N\Bigl(Ax_{N-1} + B u_{N-1}\Bigr) \biggr] =: \mathcal{M}_{N-1}(x_{N-1}),
\end{align}
which maps the state \(x_{N-1}\) to the optimal cost at time \({N-1}\). 
Moving one step back to \(k=N-2\), requires solving
\begin{align}\label{eq:def_quad2}
\min_{u_{N-2}}\mathbb{E}_{w_{N-2}} &\!\Biggl[r(u_{N-2}) \!+\!\! D_{q_{N-1}}\Bigl(Ax_{N-2}\! \!+\!\! B u_{N-2},\!-w_{N-2} \Bigr) \nonumber \\ &+ \mathcal{M}_{N-1}\Bigl(Ax_{N-2}\!+\!Bu_{N-2}\!+\!w_{N-2}\Bigr)\Biggr]
\end{align}
\vspace{-3mm}

\subsubsection*{Quadratic Cost-to-Go: Motivation and Advantages}

The expectation in \eqref{eq:def_quad2} is tractable if we assume a \emph{quadratic cost-to-go} of the form
\[
\mathcal{M}_k(x_k)=x_k^\top M_k x_k,
\]
with \(M_k\succeq 0\). This choice offers a major advantage: when the cost-to-go is quadratic, the expectation over the disturbance \(w_{N-2}\) can be computed analytically via a \emph{completion-of-squares} argument. Indeed, substituting the quadratic cost-to-go into \eqref{eq:def_quad2} yields
\begin{align}\label{eq:def_quad3}
\min_{u_{N-2}} \!\E_{w_{N-2}}\!\! \Biggl[& r_{N\!-2}(u_{N\!-2}) \!\!+\!\! D_{q_{N-1}}\!\Bigl(\!Ax_{N\!-2}\!\!+\!\!Bu_{N-2},\!-w_{N\!-2}\!\Bigr) \nonumber\\[1mm]
&+\; \Bigl(Ax_{N-2}+Bu_{N-2}+w_{N-2}\Bigr)^\top M_{N-1} \nonumber \\&\cdot\Bigl(Ax_{N-2}+Bu_{N-2}+w_{N-2}\Bigr) \Biggr].
\end{align}
The expectation of the quadratic in \eqref{eq:def_quad3} can be split into a deterministic term and a constant,
    \begin{align*}
        \E_{w_{N-2}} \Bigl[ &\Bigl(Ax_{N-2}+Bu_{N-2}+w_{N-2}\Bigr)^\top M_{N-1} \\&\cdot\Bigl(Ax_{N-2}+Bu_{N-2}+w_{N-2}\Bigr)\Bigr] \\
        = \Bigl(Ax_{N-2}&+Bu_{N-2}\Bigr)^\top M_{N-1} \Bigl(Ax_{N-2}+Bu_{N-2}\Bigr)\\& + \E_{w_{N-2}}[w_{N-2}^\top M_{N-1} w_{N-2}],
    \end{align*}
since \(\E_{w_{N-2}}[w_{N-2}]=0\) by Assumption \ref{asum:w}. And it has been already established that 
\begin{align*}
    &\mathbb{E}_{w_{N-2}} D_{q_{N-1}}\Bigl(Ax_{N-2} \!+\! B u_{N-2},-w_{N-2} \Bigr) \\&=q_{N-1}\Bigl(Ax_{N-2} \!+\! B u_{N-2} \Bigr)+c_{N-2}, 
\end{align*}
where $c_{N-2} =\E_{w_{N-2}}\!\Bigl[ D_{q_{N-1}}\Bigl(0, -w_{N-2}\Bigr) \Bigr].$

Thus, the optimization at $k=N-2$ in \eqref{eq:def_quad3} reduces to solving the following deterministic optimization:
\begin{align}\label{eq:def_quad4}
\min_{u_{N-2}} \Biggl[&r_{N-2}(u_{N-2}) \!+ \!q_{N-1}(Ax_{N-2} \!+\! B u_{N-2} ) \nonumber\\ &+ \!\!\Bigl(Ax_{N-2}\!+\!\!Bu_{N-2}\Bigr)^\top \!M_{N\!-1} \!\Bigl(\!Ax_{N-2}\!+\!Bu_{N-2}\!\Bigr)\!\Biggr].
\end{align}

Now, to  recover the recursion in \eqref{eq:def_quad_I} and to preserve dynamic programming tractability, we define

\begin{align}\label{eq:cost_to_go}
  &p_{N-1}\Bigl(Ax_{N-2} + B u_{N-2} \Bigr)\!:=\!q_{N-1}\Bigl(Ax_{N-2} \!+\! B u_{N-2} \Bigr) \nonumber \\ &+\Bigl(Ax_{N-2}\!+\!Bu_{N-2}\Bigr)^\top \!M_{N-1}\Bigl(Ax_{N-2}+Bu_{N-2}\Bigr), 
\end{align}  
where 
\eqref{eq:cost_to_go} will act as a backward Riccati-like equation.
\vspace{3mm}

Using the condition \eqref{eq:cost_to_go}, we arrive at the following optimization problem:
\begin{align}\label{eq:def_quad33}
 &J_{N-2}(x_{N-2})= D_{q_{N-2}}\Bigl(Ax_{N-3} \!+\! Bu_{N-3},\! -w_{N-3}\Bigr)  \!\nonumber\\&+\!\min_{u_{N-2}} \Biggl[r_{N-2}(u_{N-2}) \!+ \!p_{N-1}\Bigl(Ax_{N-2} \!+\! B u_{N-2} \Bigr) \Biggr] \!+\! c_{N\!-\!2},
\end{align}
which corresponds to the problem in \eqref{eq:DP_simplified} for $k=N-2$. We have thus performed one complete step of the DP recursion.

\subsubsection{Derivation of the Optimal Controller}

Once the DP recursion is established, we solve
\begin{equation*}
\min_{u_k} \Bigl[r_k(u_k) + p_{k+1}(Ax_k+Bu_k)\Bigr] = x_k^\top M_k x_k
\end{equation*}
via the Karush-Kuhn-Tucker (KKT) conditions, which are:
\begin{align}
    &\nabla r_k(u_k^\ast) + B^\top \nabla p_{k+1}(Ax_k+Bu_k^\ast) = 0,\label{eq:hi11}\\[1mm]
   & A^\top \nabla p_{k+1}(Ax_k+Bu_k^\ast) = 2 M_k x_k. \label{eq:hi22}
\end{align}
From \eqref{eq:hi22}, we have: $\nabla p_{k+1}(Ax_k+Bu_k^\ast) = 2 A^{-\top} M_k x_k$. Replacing this expression in \eqref{eq:hi11} gives the optimal control law:
\begin{equation}\label{eq:optimal_controller1}
    u_k^\ast = \nabla r_k^\ast \Bigl(-2B^\top A^{-\top}  M_k x_k\Bigr),
\end{equation}
where \(\nabla r_k^\ast\) is the gradient of the Fenchel dual of \(r_k\).
Thus, the Bregman divergence framework yields a \emph{nonlinear} state feedback law.

Here, we used the finite-horizon problem to motivate our infinite-horizon problem, the one we intend to focus on in this paper.

\subsection{The Infinite Horizon Problem}
We use the same approach used for the finite horizon, but formally, we adapt it to the infinite horizon as follows.
We employ Bellman's Optimality Principle. The value function \( J(x_k) \) represents the minimum expected cost starting from the state \( x_k \) at time step \( k \):
\begin{align}
    J(x_k) = \min_{u_k} \mathbb{E}_{w_k} \biggl[& D_q\Bigl(Ax_{k-1}+Bu_{k-1},-w_{k-1}\Bigr) + r(u_k) \nonumber \\ &+ J(x_{k+1}) \biggr]
    \label{eq:bellman_infinite_bregman}
\end{align}
We assume the following form of the value function,
\begin{align}
    J(x_k) &= x_k^\top M x_k +D_q\left(Ax_{k-1}+Bu_{k-1},-w_{k-1}\right)+ c_k
    \label{eq:value_function_bregman}
\end{align}
where \( M\succeq 0\) and $c_k$ is a constant to be defined. 
We substitute \eqref{eq:value_function_bregman} into \eqref{eq:bellman_infinite_bregman} to get:
\begin{align*}
   \min_{u_k} \mathbb{E}_{w_k} &\biggl[D_q\Bigl(Ax_{k-1}+Bu_{k-1},-w_{k-1}\Bigr) + r(u_k)\\ &+ \Bigl(Ax_k+Bu_k+w_k\Bigr)^\top M \Bigl(Ax_k+Bu_k+w_k\Bigr)\\
   &+D_q\Bigl(Ax_{k}+Bu_{k},-w_{k}\Bigr)+ c_{k+1}\biggr]  \\ &=x_k^\top M x_k +D_q\Bigl(Ax_{k-1}+Bu_{k-1},-w_{k-1}\Bigr)+ c_k
\end{align*}
which is equivalent to
\begin{align}\label{eq:pre-final}
   \min_{u_k} \mathbb{E}_{w_k} \biggl[ &r(u_k) +D_q\Bigl(Ax_{k}+Bu_{k},-w_{k}\Bigr)+\!c_{k+1}\nonumber\\&+\Bigl(Ax_k+Bu_k+w_k\Bigr)^\top\! M \Bigl(Ax_k\!+\!Bu_k\!+\!w_k\Bigr)\!\biggr] \nonumber  \\&=x_k^\top M x_k+ c_k
\end{align}
Now, we need to expand the left-hand side and take expectations. Again, similarly to the finite horizon case; using Lemma \ref{lem:expectation_property}, along with Assumption \ref{asum:q}, the expectation term, $\mathbb{E}_{w_k} D_q\left(Ax_{k}+Bu_{k},-w_{k}\right)$, in \eqref{eq:pre-final} can be written as  
\begin{align}\label{eq:bregExpect}
    &\mathbb{E}_{w_{k}} \biggl[ D_q\Bigl(Ax_{k} + B u_{k}, -w_{k}\Bigr) \biggr] = q\Bigl(Ax_{k} + B u_{k}\Bigr)  +d_{k}
\end{align}
where $d_{k}=  \mathbb{E}_{w_{k}} w_{k} ^\top  \nabla q(-w_{k})- \mathbb{E}_{w_{k}} q(-w_{k}) $, which is independent of $x_{k}$ and $u_{k}$. 
Replacing \eqref{eq:bregExpect} in \eqref{eq:pre-final}, and taking the expectation while noting that $\mathbb{E} \left[w_k\right]=0$, we get: 
\begin{align*}
   \min_{u_k}  \biggl[& r(u_k) + \Bigl(Ax_k+Bu_k\Bigr)^\top M \Bigl(Ax_k+Bu_k\Bigr)\\&+q\Bigl(Ax_{k} + B u_{k}\Bigr)\biggr] 
   +c_{k+1}+ d_{k}+ \mathbb{E}_{w_k} w_k^\top M w_k\\& =x_k^\top M x_k + c_k
\end{align*}

Defining $c_k:=\sum_{j=k}^{\infty}\left(d_j+\mathbb{E}\left[{w_j}^\top Mw_j\right]\right)$, we get \( c_k = c_{k+1}+d_{k}+ \mathbb{E}_{w_k} w_k^\top M w_k\), so the constant terms cancel out. 

\vspace{0.4em} The resulting \emph{deterministic optimization} is: 
\begin{align}\label{eq:final}
   \min_{u_k}  \biggl[ &r(u_k) + \Bigl(Ax_k+Bu_k\Bigr)^\top M \Bigl(Ax_k+Bu_k\Bigr)\nonumber \\&+q\Bigl(Ax_{k} + B u_{k}\Bigr)\biggr] =x_k^\top M x_k
\end{align}

Defining the convex, even and positive function $p(\cdot)$ to be:
\begin{equation}\label{eq:pqHi}
    p(x_k):= q(x_k)+x_k^\top Mx_k
\end{equation}
allows to rewrite \eqref{eq:final} as 
\begin{align}\label{eq:final2}
   &\min_{u_k}  \biggl[ r(u_k) +p\Bigl(Ax_{k} + B u_{k}\Bigr)\biggr] =x_k^\top M x_k
\end{align}
Then, the Karush-Kuhn-Tucker (KKT) conditions are:
\begin{align}
    &\nabla r(u_k^\ast) + B^\top \nabla p(Ax_k+Bu_k^\ast) = 0,\label{eq:hi1}\\[1mm]
   & A^\top \nabla p(Ax_k+Bu_k^\ast) = 2 M x_k. \label{eq:hi2}
\end{align}
which yield the optimal controller $u_k^\ast = \nabla r^\ast \Bigl(-2B^\top A^{-\top}  M x_k\Bigr).$
 Moreover, a Riccati-like equation for \(M\) emerges from algebraic manipulating of the KKT conditions:
\begin{align}\label{eq:pqr12}
    &p^\ast(\xi) + r^\ast(-B^\top \xi) = \frac{1}{4}\,\xi^\top A M^{-1} A^\top \xi,\\
    &\text{with } p(x)=q(x)+x^\top M x
\end{align}
\textbf{Stability via a Lyapunov Function:} Finally, the function \(p(\cdot)\) defined in \eqref{eq:pqHi} serves as a Lyapunov function proving stability. Indeed, by evaluating
\[
p(Ax_k+Bu_k^\ast) - p(x_k),
\]
and using equations \eqref{eq:def_quad_I} and \eqref{eq:pqHi}, we have:
\begin{align*}
    p(Ax_k+Bu_k^\ast)-p(x_k)&=x_k^\top M x_k \!-\! r(u^\ast)\! \!- \!\!(q(x)\!+\!x^\top Mx)\\&=-r(u^\ast)-q(x)
\end{align*}
which is negative.
Thus, we summarize our findings in the following theorem in which we derive the optimal controller.

\begin{theorem}[Optimal Controller Synthesis via KKT Conditions]
\label{thm:controller_synthesis}
Let Assumptions \ref{asum:w} and \ref{asum:AB} hold. Let $M$ be a positive definite matrix, and \( q(\cdot),p(\cdot),r(\cdot) \) be strictly convex, even, and non negative functions satisfying the Riccati-like equations:
\begin{align}
    &p^\ast(\xi)+ r^\ast \Bigl( B^\top  \xi\Bigr)=\frac{1}{4}\xi^\top AM^{-1}A^\top \xi\label{eq:pqr1}\quad \forall \xi\\
    &p(x)=q(x)+x^\top Mx \quad \forall x.\label{eq:pqr2}
\end{align}
along with Assumption \ref{asum:q}. 
Then, the optimal control law, which solves
\begin{align}\label{eq:breg-cost-again}
\min_{\{u_k\}} &\lim_{N\rightarrow \infty}\mathbb{E}_{\{w_k\}} \frac{1}{N}   \biggl[ D_r(u_0, 0)\\ &+\sum_{k=1}^{N-1}  D_q\Bigl(Ax_{k-1} + Bu_{k-1}, -w_{k-1}\Bigr) \nonumber +  D_r(u_k,0)\biggr] 
\end{align}\emph{\textcolor{black}{stabilizes the system}}, and is explicitly given by
\begin{equation}\label{eq:optimal_controller}
    u_k^\ast = \nabla r^\ast \left( -2 B^\top  A^{-\top}  M x_k \right).
\end{equation}
\end{theorem}

In the theorem above, the stability follows from using $J(x_k)$ in \eqref{eq:value_function_bregman} as a Lyapunov function.


\subsection{Summary of the Bregman Divergence Framework}
In summary, by adopting a Bregman divergence framework and imposing a quadratic structure on the cost-to-go, we decouple the stochastic and deterministic components of the dynamics. This decoupling not only simplifies the dynamic programming recursion but also yields a controller with inherent stability properties. The resulting framework generalizes the classical LQR approach, enabling the design of more versatile control strategies.

\subsection{Discussion on the Controller Synthesis}
\textbf{Theorem \ref{thm:controller_synthesis}} provides an explicit formulation for an optimal nonlinear state-feedback controller applicable to linear systems when minimizing non-quadratic cost functions, while simultaneously ensuring system stability. This development is highly valuable for practical implementations, as it significantly expands the repertoire of viable controller designs beyond traditional quadratic frameworks.

In practice, selecting state and control cost functions must adhere to the constraints specified in Theorem~\ref{thm:controller_synthesis}. In particular, the designer cannot independently choose arbitrary functions \(q(\cdot)\) and \(r(\cdot)\), because the theorem’s requirement for a quadratic cost-to-go forces a dependency between them. As a result, one must be selected a priori---either \(q\) or \(r\)---with the other being determined by the Riccati-like equation \eqref{eq:pqr1}. The following section details how to meet these requirements through two principal design strategies:

\begin{enumerate}

    \item \textbf{State Cost Specification:} The designer initially selects a state cost function \(q(\cdot)\). Then, for a given positive definite matrix \(M\), the corresponding control cost function \(r(\cdot)\) is derived via Equations~\eqref{eq:pqr1} and \eqref{eq:pqr2}. However, \(M\) must be chosen carefully to ensure that the resulting \(r(\cdot)\) is both convex and positive. Thus, the design problem reduces to finding an \(M \succ 0\) that satisfies these criteria, a task that can be addressed using convex optimization techniques.

 \item \textbf{Control Cost Specification:} Alternatively, the designer may begin by choosing a control cost function \(r(\cdot)\). With a selected positive definite matrix \(M\), the corresponding state cost \(q(\cdot)\) is computed using the same equations. In this case, the challenge is to find an \(M \succ 0\) that ensures the derived \(q(\cdot)\) is convex and positive, which can similarly be achieved via convex optimization.


\end{enumerate}

This flexible framework facilitates the design of advanced control strategies, including a principled way of safety control, bang-bang control, and sparse control. Consequently, Theorem \ref{thm:controller_synthesis} opens up a new landscape of possibilities in control engineering, enabling the development of more tailored and efficient controllers for a wide range of applications.
\section{Designing the Cost Functions}
\label{sec:cost_design}

For reasons of space, we only consider the cases where the designer choses state cost $q(\cdot)$ or controller cost $r(\cdot)$. 
\subsection{Selecting the State Cost \(q\)}

Here, the designer selects an even, positive, and strictly convex state cost function \(q(\cdot)\). 
For fully or over-actuated systems, Theorem~\ref{thm:chooseQ} establishes necessary and sufficient conditions for a positive definite matrix \(M\) to exist such that $r(\cdot)$ is convex and positive. This guarantees the existence of the optimal stabilizing controller as stated in Equation~\eqref{eq:optimal_controller}.




\begin{theorem}\label{thm:chooseQ}
Consider a fully actuated or overactuated system, i.e., \(B\in \mathbb{R}^{n\times m}\) with \(m\geq n\). Let \(q: \mathbb{R}^n \to \mathbb{R}\) be a strictly convex, even, and positive function chosen by the designer. Then, for \(M\succ 0\), the function \(r\) given by
\begin{align}
   &\tilde r^\ast \left( \xi \right) = -p^\ast\big((B^\top)^\dagger\xi\big) + \frac{1}{4} \xi^\top B^\dagger A M^{-1} A^\top (B^\top)^\dagger \xi, \label{eq:pqr111}\\[1mm]
   &\text{with } p(x) = q(x) + x^\top M x \label{eq:pqr222},\quad \text{ and }\quad r(\cdot):= \tilde{r}^{\ast\ast}(\cdot)
\end{align}
is convex and positive if and only if \(M\) satisfies
\begin{align}
   M-\frac{1}{2}A^{T}MA \preceq \frac{1}{2}A^{T}\nabla^2q(x)A, \quad \forall x. \label{eq:rConvex}
\end{align}
\end{theorem}

 The proof idea is in the Appendix. 
 
 Theorem~\ref{thm:chooseQ} thus reduces the design problem to finding a positive definite matrix \(M\) that satisfies \eqref{eq:rConvex}. Once \( M \) is obtained, the corresponding cost function \( r \) is computed via \eqref{eq:pqr111}.
Moreover, one may incorporate alternative formulations that choose $M$ from the feasibility set \eqref{eq:rConvex}. Such modifications can influence controller performance, as demonstrated in the simulation results. 

Note that we have extended Theorem \ref{thm:chooseQ} to address the more challenging problem of state‐cost design for underactuated systems; this result will appear in the next revision of the manuscript.

\subsection{Selecting the Control Cost \(r\)}

Here, the designer selects an even, positive, and strictly convex control cost function \(r(\cdot)\). 
Theorem~\ref{thm:chooseR} establishes necessary and sufficient conditions for a positive definite matrix \(M\) to exist such that $q(\cdot)$ is convex and positive. This guarantees the existence of the optimal stabilizing controller as stated in Equation~\eqref{eq:optimal_controller}.

\begin{theorem}\label{thm:chooseR} 
      Let \( r: \mathbb{R}^n \to \mathbb{R} \) be a convex, even, and positive function chosen by the designer. Consider a stable matrix \( A \). Then, for a given \(M\succ 0\), a causal time-invariant control policy that solves Problem \ref{pb:bregman} exists, i.e., the functions \(p(\cdot)\) and $q(\cdot)$ given by
\begin{align}
    p^\ast(\xi) \!&=\! -r^\ast \left( B^\top \xi \right) \!+\! \frac{1}{4} \xi^\top A M^{-1} A^\top \xi\!,\label{eq:UGH}  \\
    q(x) &= p(x) - x^\top M x \label{eq:UGH2}
\end{align}
are convex and positive, if and only if \(M^{-1}\) satisfies the condition:   
    \begin{align}
     &\frac{1}{2} \left( A M^{-1} A^\top - M^{-1} \right) 
        \!\preceq \!B \nabla^2 r^\ast\left( B^\top \xi \right) B^\top 
        \!\preceq\! \frac{1}{2} A M^{-1} A^\top 
        \label{eq:pqConvex}
    \end{align}
  Furthermore, suppose that $r^\ast(\cdot)$ is strongly convex and smooth satisfying
    \begin{equation}\label{eq:rCvxSmooth}
        \xi^\top L_r \xi \le r^\ast\bigl(B^\top \xi\bigr) \le \xi^\top U_r\xi \quad \forall \xi \in \mathbb{R}^n,
    \end{equation}
    for some positive definite matrices \( L_r,U_r \succ 0 \). Then, for any matrix \( A \), for a given \(M\succ 0\), a time-invariant control policy that solves Problem \ref{pb:bregman} exists, if $M^{-1}$ satisfies the following convex feasibility condition:
    \begin{equation}\label{eq:cvxopt}
    \begin{aligned}
                U \preceq \frac{1}{4} A M^{-1} A^\top \preceq L + M^{-1}.
    \end{aligned}           
    \end{equation}
\end{theorem}
The proof idea is in the Appendix.

In the case where $A$ is stable, the designer enjoys greater flexibility in selecting the cost function $r$ which need not be strongly convex. Theorem~\ref{thm:chooseR} reduces the design problem to finding a positive definite matrix \(M^{-1}\) that satisfies the convex condition \eqref{eq:pqConvex} (which can be further simplified by taking the infemum and supremum over all $\xi$ for $\nabla^2r^\ast(B^\top \xi).$)

In the case where $A$ is any matrix (stable or unstable), $r$ is chosen to be any strongly convex and smooth, and thus Theorem~\ref{thm:chooseR} reduces the design problem to finding a positive definite matrix \(M\) that satisfies the convex program \eqref{eq:cvxopt}. Once \( M \) is obtained, the corresponding cost functions \( q(\cdot)\) and $p(\cdot)$ are computed from \eqref{eq:UGH}, \eqref{eq:UGH2}. 
\section{Applications and Simulations}
For simplicity, consider a system with scalar state. We present a bang-bang controller, and an elastic-net controller.
\subsection{Bang-Bang Controller}
\begin{figure}[h]
  \centering
  \subfloat[States]{%
    \includegraphics[width=0.7\columnwidth]{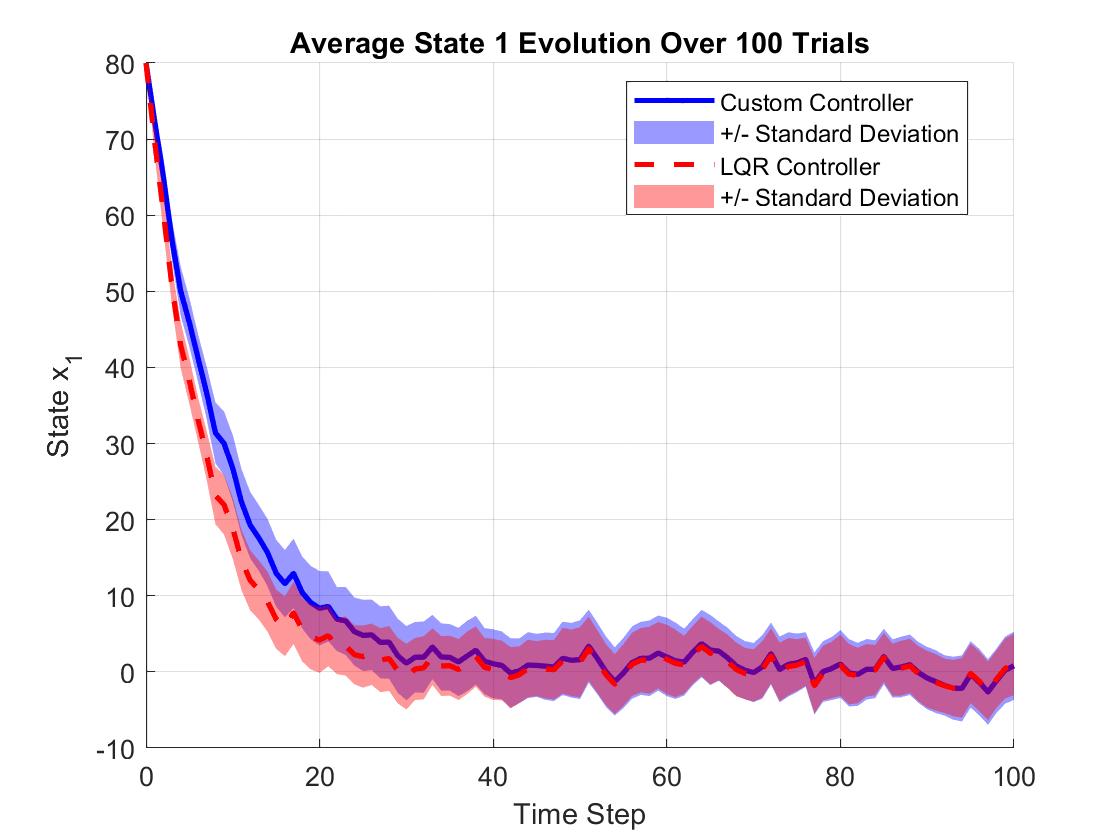}%
  }\hfill
  \subfloat[Control inputs]{%
    \includegraphics[width=0.7\columnwidth]{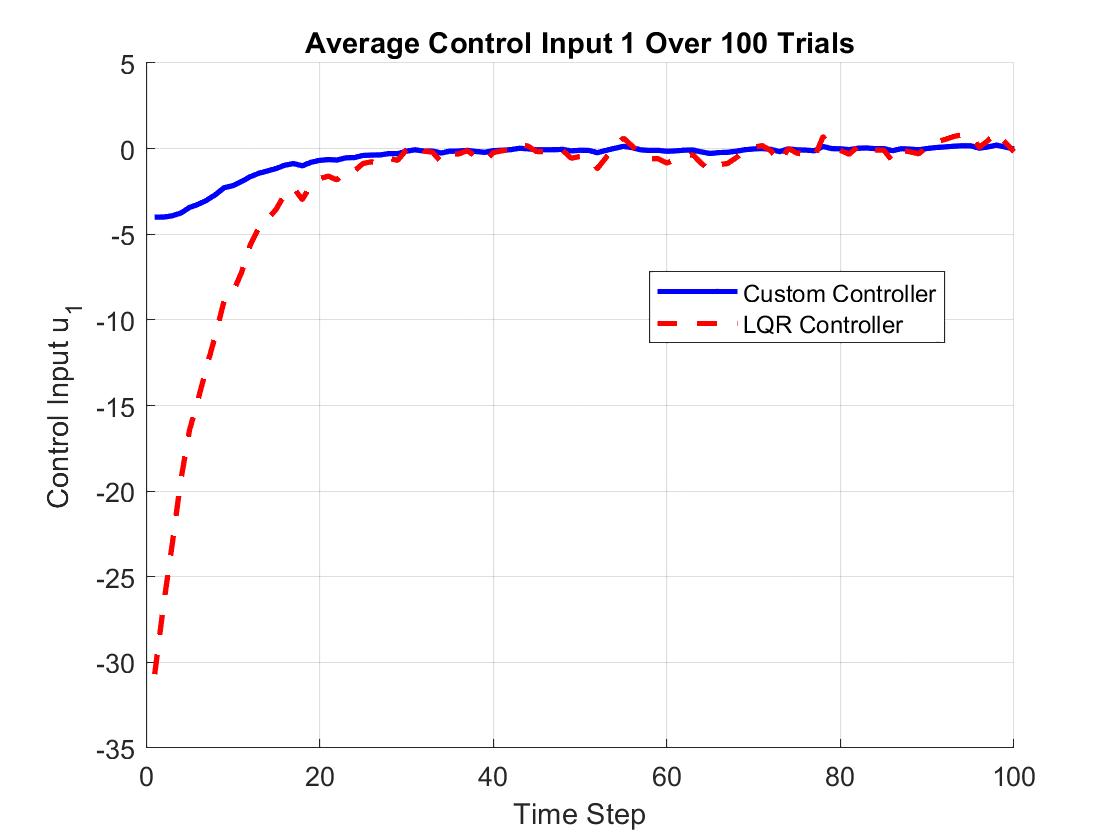}%
  }\\[1ex]
  \subfloat[Cost functions]{%
    \includegraphics[width=0.7\columnwidth]{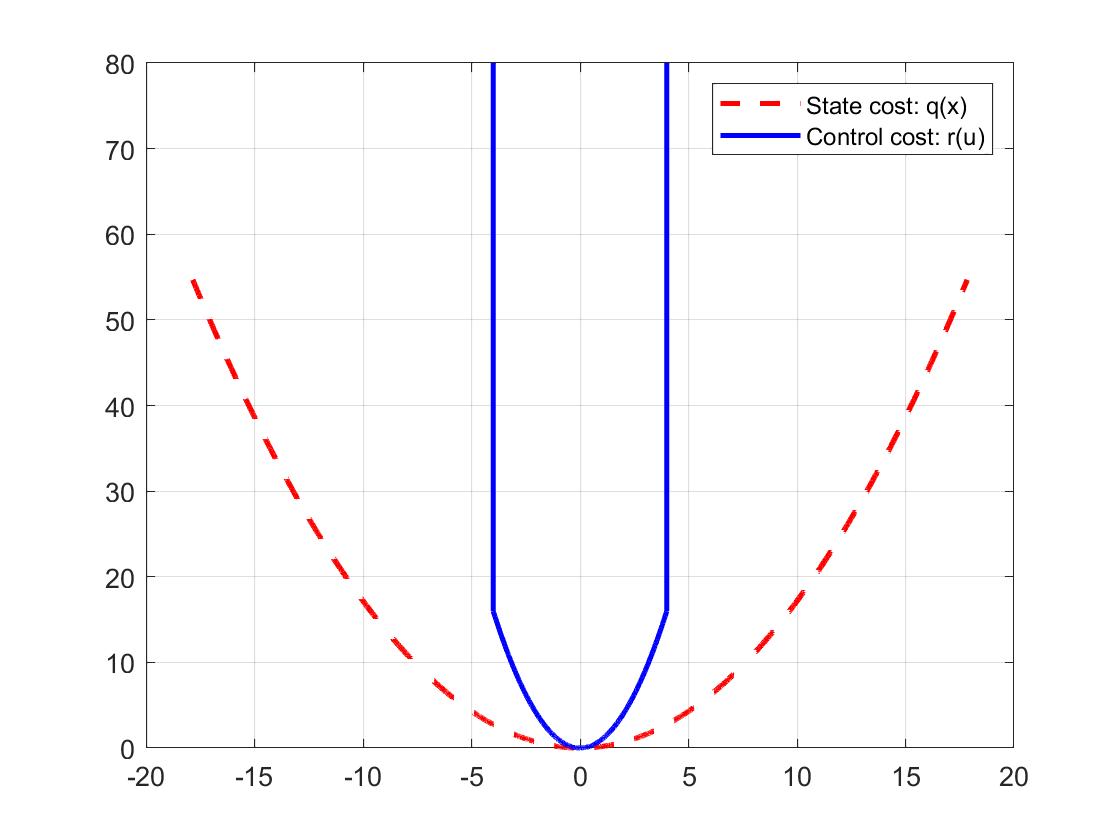}%
  }
  \caption{State trajectories, control inputs and costs with the bang-bang control cost, for a scalar system with $t=4$.}
  \label{fig:bangR-scalar}
\end{figure}

It is often the case that in practice one has a budget for the amount of input control at each time step. Here we focus on $l_\infty$ constraint on the controller. In the scalar setting, we impose the following cost on the controller.
\begin{align}
r(u) &= 
\begin{cases}
u^2 & \text{if } |u| < t\\
  \infty & \text{if } |u| \geq t
\end{cases} 
\end{align}
The feasible cost on the state is then
\begin{align}
q(x) &= \begin{cases}
m(\frac{1}{a^2-b^2m}-1)x^2,\quad \text{if } |x| \leq \frac{a^2t}{mb}-tb\\
   m(\frac{1}{a^2}-1)x^2+\frac{2mbt}{a^2}|x|+t^2(\frac{mb^2}{a^2}-1),\\
   \text{if } |x| > \frac{a^2t}{mb}-tb
\end{cases}
\end{align}
The optimal controller is
\begin{align*}
    u^{\ast}(x) =
    \begin{cases}
    -\,\dfrac{b m}{a} \, x, & \text{if } \left| x \right| \leq \dfrac{a t}{b m}, \\[10pt]
    -\,a \cdot \text{sign}(x), & \text{if } \left| x \right| > \dfrac{a t}{b m}.
    \end{cases}
\end{align*}
Figure \ref{fig:bangR-scalar} shows simulations of a scalar system with the following choice of parameters:
$a=0.9,\quad b=0.1 ,\quad  m=0.7, \quad t=4$. 
It can be seen that the controller is constrained to have values less than $t$.

\subsection{Exponential Controller}
To explore how little control effort can still stabilize the system, the designer can penalize large inputs by choosing
\begin{align*}
    r(u) &= e^{|u| } - |u| - 1,
\end{align*}which grows exponentially for large $|u|$. The closed-form expression for the controller is:
\begin{align*}
    u{^\ast}(x) &=- \text{sign}( x ) \cdot \log\left( | \frac{ 2 b m }{ a } x | + 1 \right)
\end{align*}
However, the state cost doesn't admit a closed form:
\begin{align*}  
    q(x)=&\max_{y} -mx^2+\xi x-\frac{a^2}{4m}\xi^2+1\\
    &-(1+b|\xi|)(-1+\log(1+b|\xi|))\\
    \text{where }\xi&=\frac{a^2}{2m}x^{\ast2}+sign(x^\ast)b \log(1+b|x^\ast|), 
\end{align*} $\text{with } x^\ast \text{is the optimal x}$.

\begin{figure}[htbp]
  \centering
  \subfloat[States]{%
    \includegraphics[width=0.7\columnwidth]{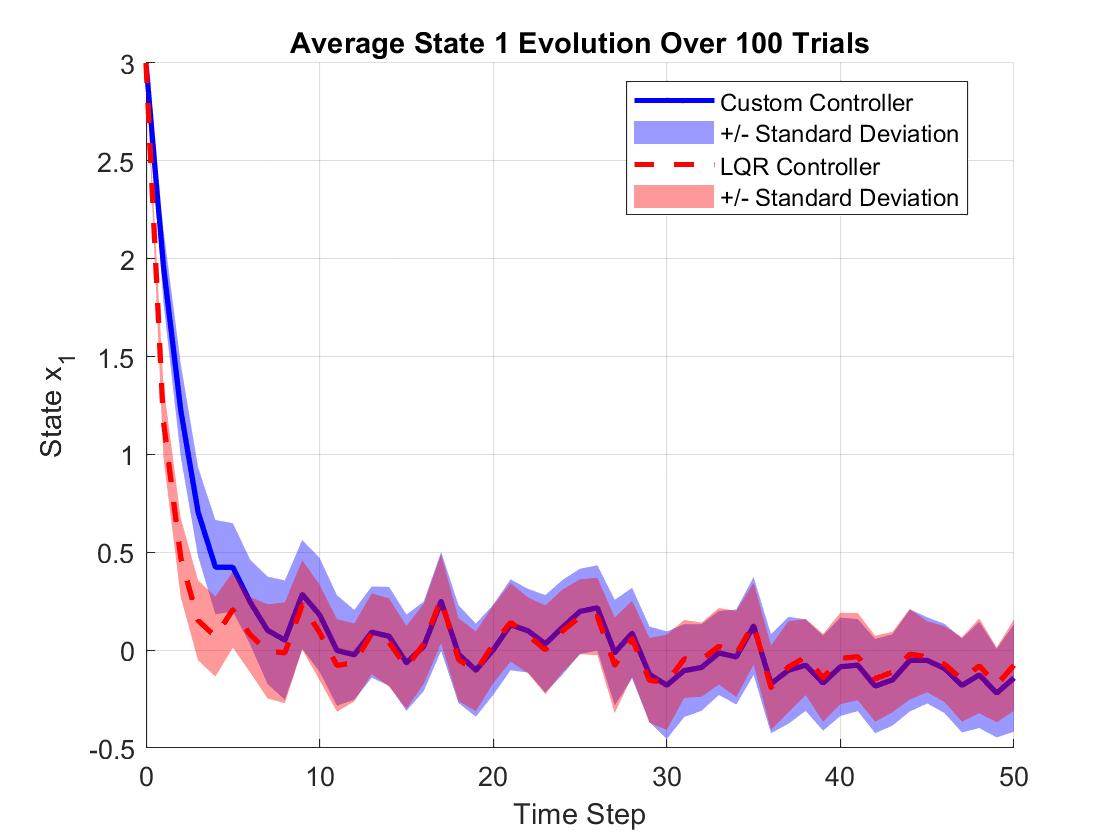}%
  }\hfill
  \subfloat[Control inputs]{%
    \includegraphics[width=0.7\columnwidth]{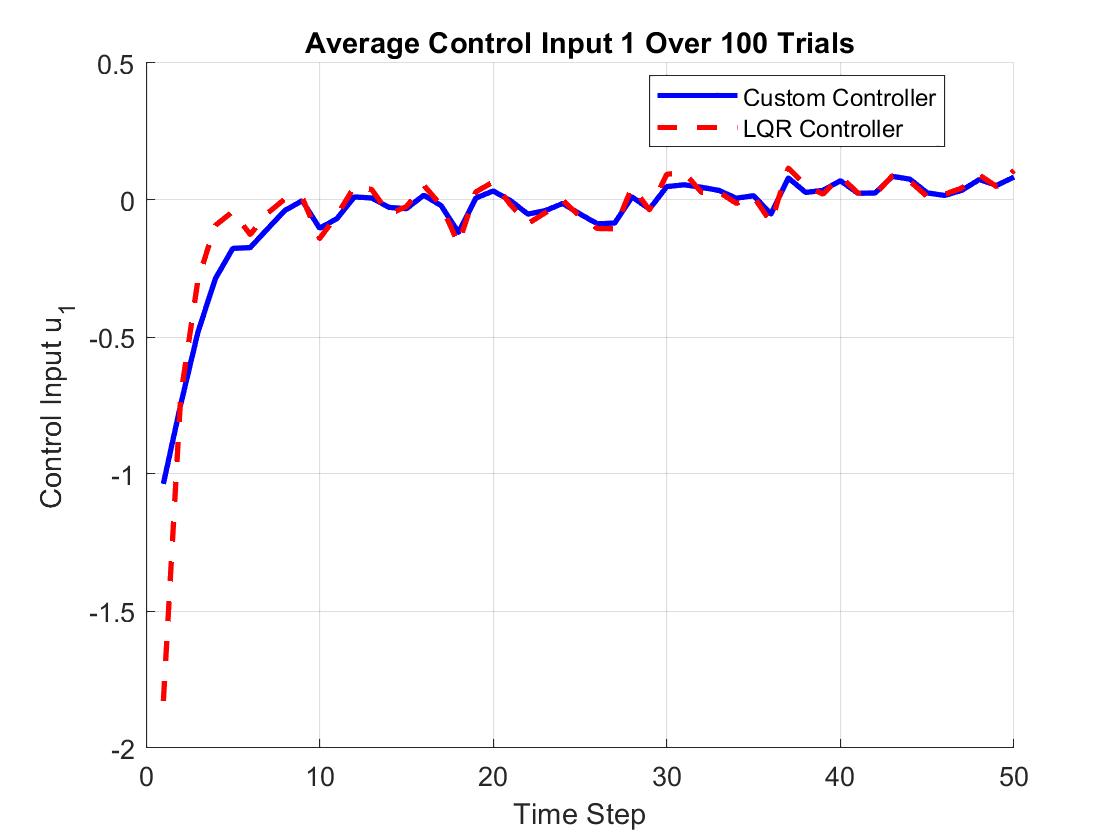}%
  }\\[1ex]
  \subfloat[Cost functions]{%
    \includegraphics[width=0.7\columnwidth]{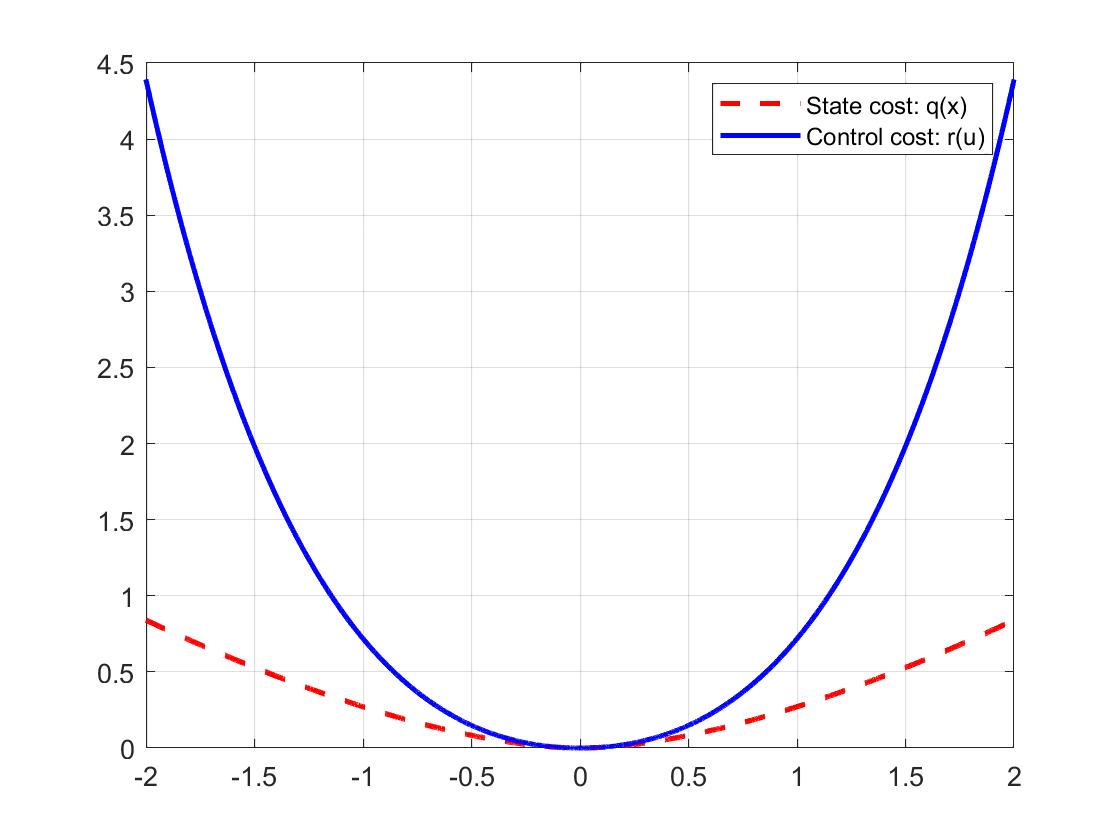}%
  }
  \caption{State trajectories, control inputs, and cost functions for an exponential control cost ($m=0.3$).}
  \label{fig:expCtrl}
\end{figure}

For our experiments, we set: ($a=0.99, b=1, $,$m=0.3$). Figure \ref{fig:expCtrl} under this controller the closed-loop uses far less input—thanks to the exponential penalty—while still keeping the state bounded. 
\subsection{Elastic-Net State Cost}

In this section, we explore an elastic-net \cite{zou2005regularization} formulation for the state cost, which is particularly attractive for enforcing sparsity in the states or controllers. Directly minimizing the \(\ell_1\)-norm would ideally promote sparsity, but its non-differentiability and lack of strong convexity can complicate controller design. By combining the \(\ell_1\) and \(\ell_2\) penalties, the elastic-net formulation ensures both sparsity and smoothness. 


Specifically, the state cost function is defined as:
\begin{align}
q(x) = |x| + \epsilon x^2,
\end{align}
where \(\epsilon > 0\) guarantees strong convexity.

Given this state cost, the feasible controller cost \(r(u)\) is derived as:
\begin{align}
    r(u) &= 
    \begin{cases}
    \frac{mb^2(\epsilon+m)}{a^2(\epsilon+m)-m}u^2  - \frac{bm}{a^2(\epsilon+m)-m}|u|  \!+ \!\frac{a^2}{4(a^2(\epsilon+m)-m)} & \\\text{if } |u| > \frac{a^2}{2mb}, \\
      \frac{mb^2}{a^2}u^2 ,\quad \text{if } |u| \leq \frac{a^2}{2mb}
    \end{cases} 
\end{align}

Accordingly, the optimal controller is expressed as:
\begin{align*}
    u^\ast(x) =
    \begin{cases}
    -\dfrac{a}{b} \, x, \quad\text{if } \left| \dfrac{2m}{a} x \right| \leq 1, \\[10pt]
    -\dfrac{a}{b} \, x + \dfrac{ \left( \left| \dfrac{2m}{a} x \right| - 1 \right) \cdot \text{sign}\left( \dfrac{2m}{a} x \right) }{ 2b (\epsilon + m) }, & \\\text{if } \left| \dfrac{2m}{a} x \right| > 1.
    \end{cases}
\end{align*}

For our experiments, we use the parameters \(a=1.2\), \(b=1\), \(\epsilon=0.01\), and two values for \(m\): a smaller value \(m=0.01\) and a larger value \(m=0.07\). The results indicate that, compared to a standard LQR controller, the elastic-net formulation drives the state to zero more rapidly, while promoting sparsity.

\begin{figure}[htbp]
  \centering
  \subfloat[States for $m=0.01$]{%
    \includegraphics[width=0.7\columnwidth]{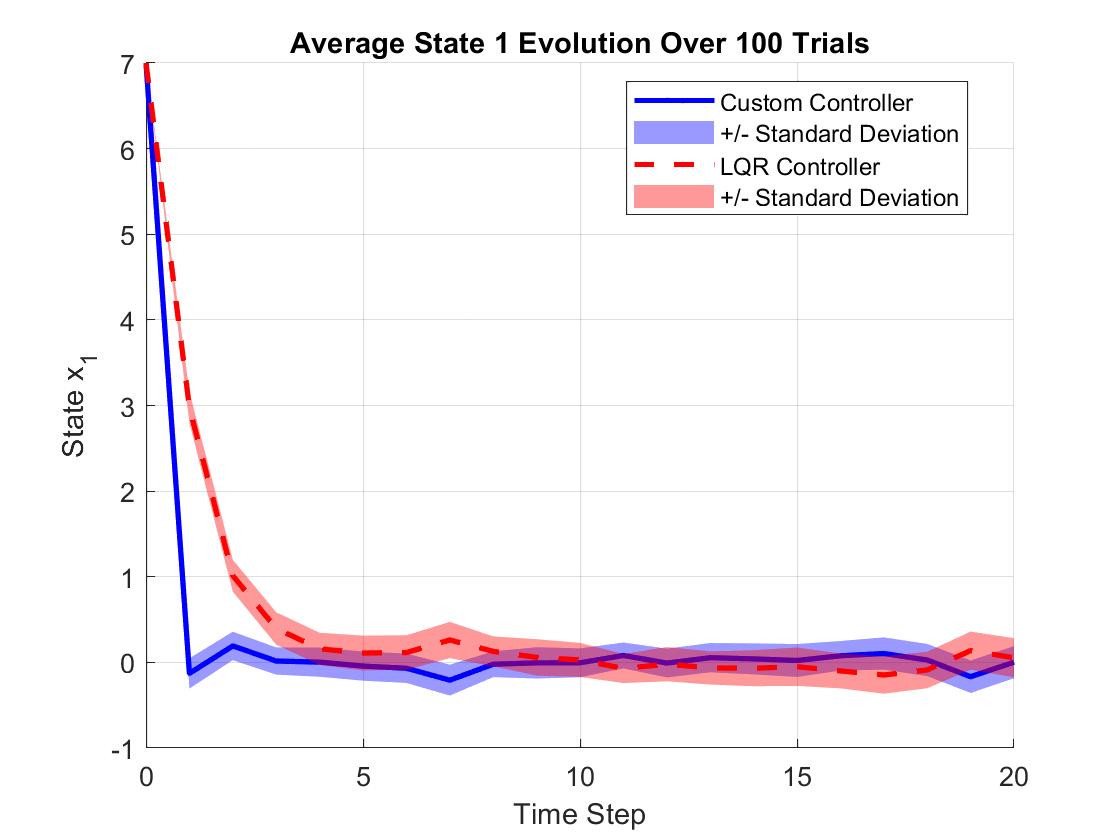}%
  }\\[1ex]
  \subfloat[Control Inputs for $m=0.01$]{%
    \includegraphics[width=0.7\columnwidth]{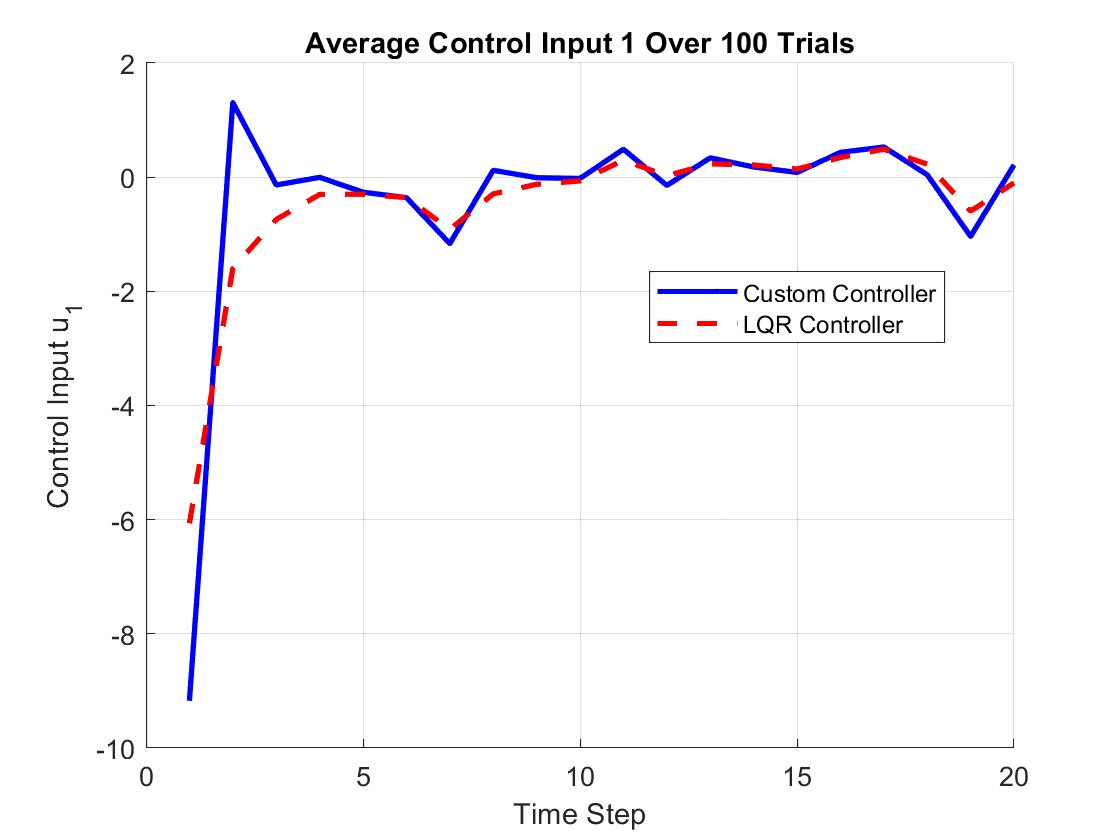}%
  }
  \caption{State trajectories and control inputs with the elastic-net state cost for $m=0.01$.}
  \label{fig:states_inputs_small}
\end{figure}

\begin{figure}[htbp]
  \centering
  \subfloat[States for $m=0.07$]{%
    \includegraphics[width=0.7\columnwidth]{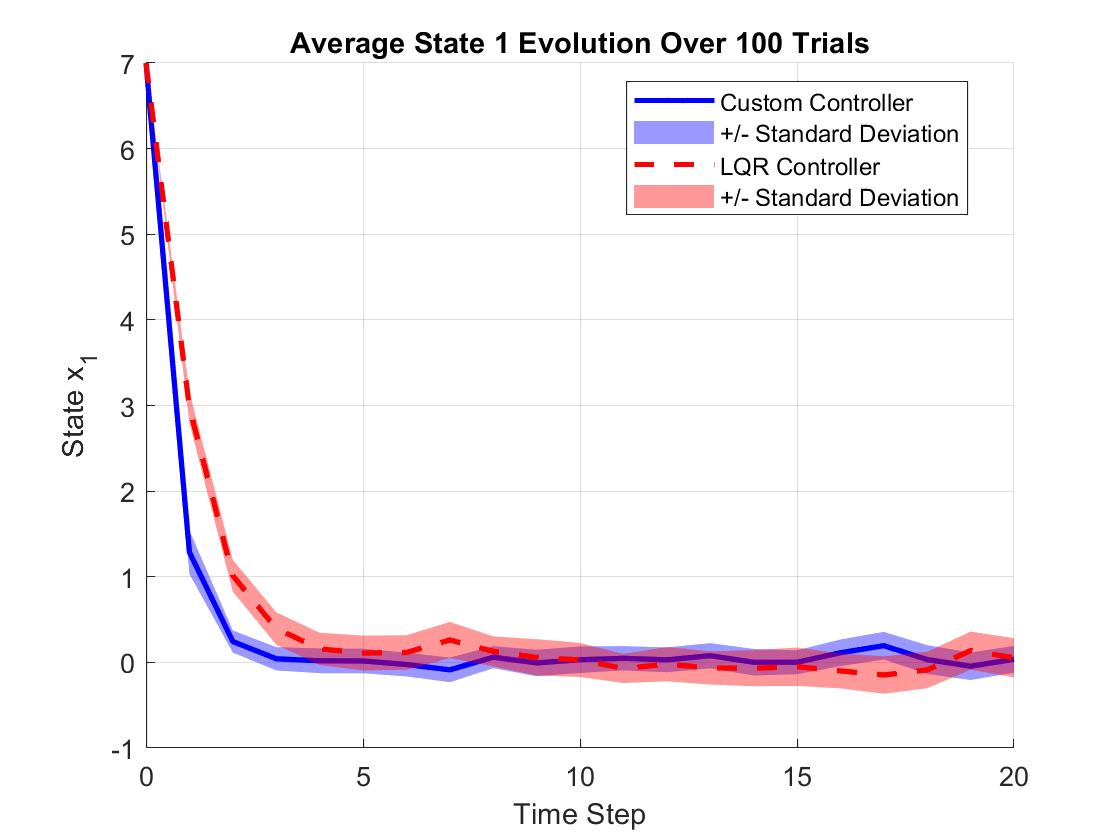}%
  }\\[1ex]
  \subfloat[Control Inputs for $m=0.07$]{%
    \includegraphics[width=0.7\columnwidth]{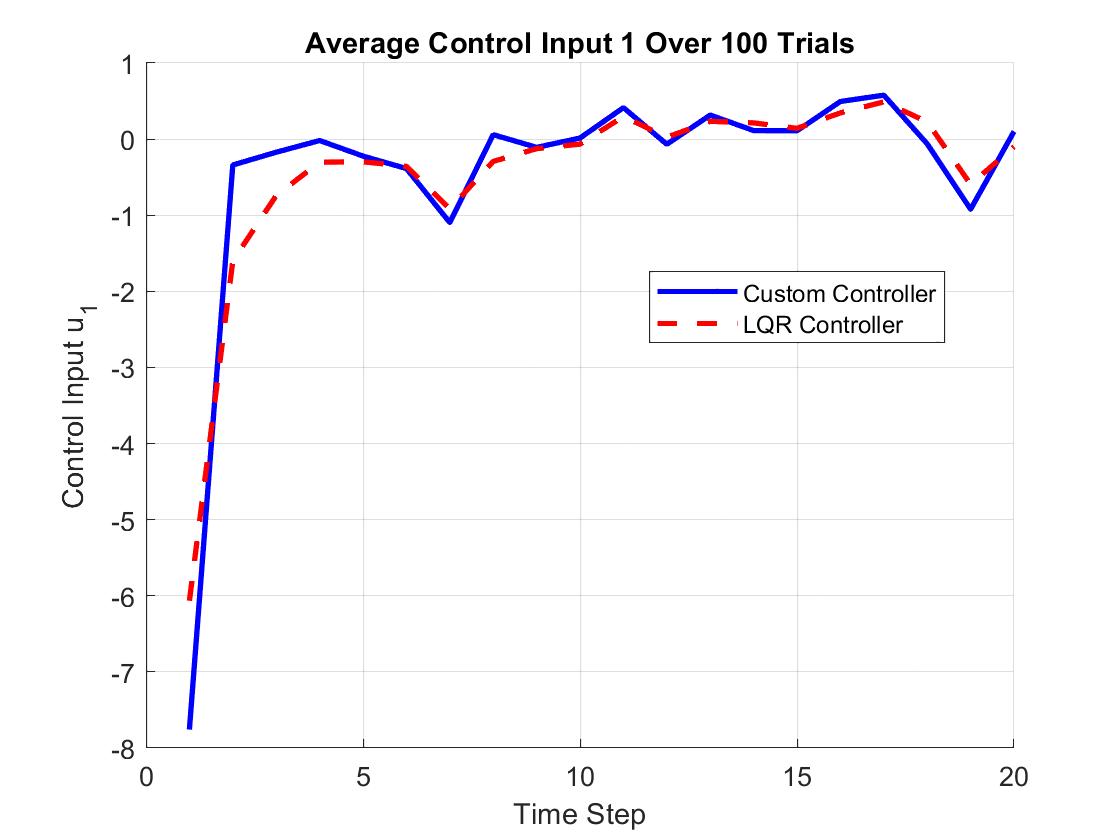}%
  }
  \caption{State trajectories and control inputs with the elastic-net state cost for $m=0.07$.}
  \label{fig:states_inputs_mid}
\end{figure}

\begin{figure}[htbp]
  \centering
  \subfloat[Elastic-net Cost Functions for $m=0.01$]{%
    \includegraphics[width=0.7\columnwidth]{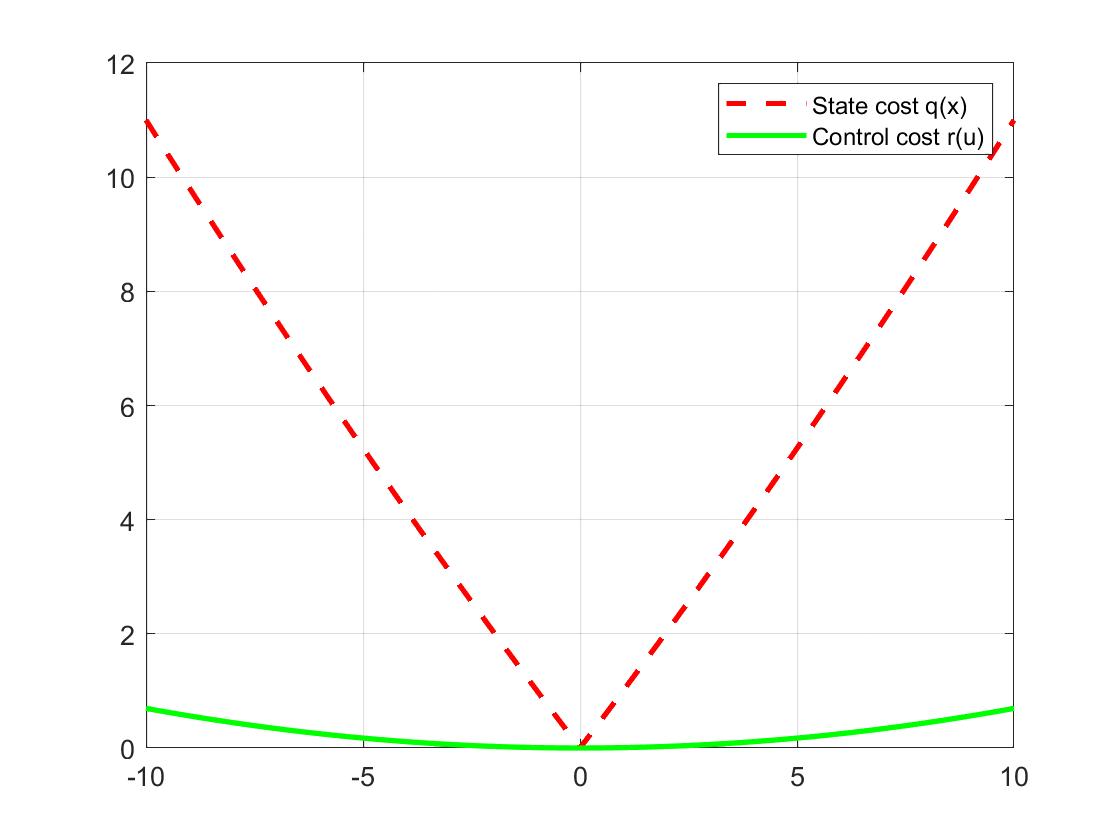}%
  }\\[1ex]
  \subfloat[Elastic-net Cost Functions for $m=0.07$]{%
    \includegraphics[width=0.7\columnwidth]{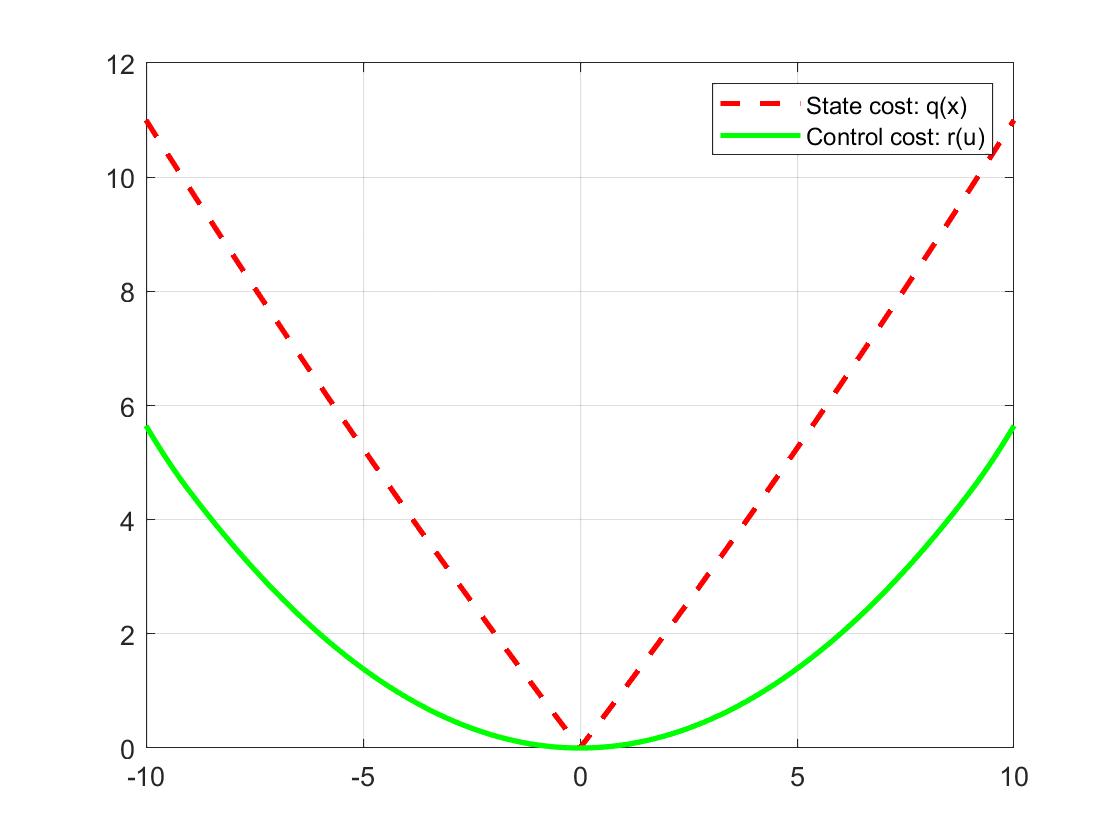}%
  }
  \caption{Comparison of the $q$ and $r$ functions for different values of $m$.}
  \label{fig:q_r_functions}
\end{figure}

Figure~\ref{fig:states_inputs_small} shows the state trajectories and control inputs for \(m=0.01\), whereas Figure~\ref{fig:states_inputs_mid} displays the corresponding results for \(m=0.07\). These figures illustrate how the choice of \(m\) influences the controller behavior. Smaller $m$ results in a more agressive controller that drives the state faster to 0. Furthermore, Figure~\ref{fig:q_r_functions} compares the \(q\) and \(r\) functions for the two different values of \(m\), highlighting the impact of the parameter on the cost function $r$ where a larger $m$ results in higher controller cost. 


\section{Conclusion}
This work has extended the classical quadratic optimal control framework by developing a novel approach based on Bregman divergence that accommodates non-quadratic convex cost functions. A key challenge addressed is the inherent difficulty introduced by disturbances. These stochastic disturbances complicate the computation of expected costs and the design of controllers, making traditional approaches intractable for infinite-horizon problems. By introducing Bregman divergence costs, and decoupling the deterministic and stochastic components in the system dynamics, our framework achieves computational tractability and facilitates the derivation of a generalized Riccati-like equation. This equation underpins the design of robust, nonlinear state-feedback controllers that guarantee system stability. The framework's effectiveness is demonstrated in an application in sparse control. The experiments underscore its potential to significantly broaden the scope of optimal control methodologies in challenging, disturbance-prone environments.
\bibliographystyle{ieeetr}
\bibliography{refs} 
\appendix
\subsection{Proof Idea of Theorems \ref{thm:chooseQ} and \ref{thm:chooseR}}
The proofs rely on the following ideas:

\begin{enumerate}
    \item \textbf{Convexity via Hessian Positivity:}  
    A function \( f \) is convex if
    \[
    \nabla^2 f(x) \succ 0 \quad \forall x.
    \]
    \item By imposing Assumption \ref{asum:q} as a constraint, convexity implies positivity of the functions.
    
    \item \textbf{Fenchel Duality and the Inverse Function Theorem:}  
    For a convex function \( f \) and its Fenchel conjugate \( f^\ast \), one obtains (via the inverse function theorem)
    \[
    \nabla^2 f\bigl(\nabla f^\ast(\xi)\bigr)\,\nabla^2 f^\ast(\xi)=I.
    \]This crucial relation shows that locally the Hessian of \( f\) is the inverse of the Hessian of \( f^\ast \).
\end{enumerate}
Combining these ideas along with some matrix inequalities provides sufficient conditions for the overall functions to be convex. 

\subsection{Proof of Theorem \ref{thm:chooseQ}}
We present the proof of Theorem \ref{thm:chooseQ} to further illustrate the argument. First we start with conditions on $p(\cdot)$. As defined in \eqref{eq:pqr222}, we have 
\begin{align*}
    p(x)=q(x)+x^\top Mx.
\end{align*}
It turns out by setting $V(x) := p(x)$, the Lyapunov function, we observe that $M\succ0$ is a sufficient condition for stability. Therefore, we will only consider the cases where $M\succ0$. Now for the conditions on $r(\cdot)$, we utilize the following result from convex analysis:
\begin{lemmma}\label{lemm:posFenchel}
    If a function $f$ is convex, positive, and $f(0)=0$, then its Fenchel dual $f^\ast$ is also convex, positive and $f^\ast(0)=0$, and vice versa. 
\end{lemmma}
In order for equation \eqref{eq:pqr1} to hold, we need to ensure $r^{\ast}(\cdot)$ is convex, even and positive. Then, using Lemma \ref{lemm:posFenchel}, we will observe that $r^{\ast\ast}$ is convex, even, positive and by choosing $r:= r^{\ast\ast}$, the results follow. We will make use of the following auxiliary result.
\begin{lemmma}\label{lemm:ineq}
    If $f^\ast(\xi)<\frac{1}{4} \xi^\top X^{-1}\xi$ $\forall \xi$, then $f(x)> x^\top Xx$ for all $x \in \mathbb{R}^n$
\end{lemmma}
We start with the conditions on the positivity of $r^\ast(\cdot)$. From \eqref{eq:pqr1}, we would need that 
\begin{align*}
    \xi^\top M\xi>p^\ast(2A^{-\top} M\xi) \quad \forall \xi.
\end{align*}
Applying a change of variable $\tilde{\xi}:= 2A^{-\top} M\xi$
\begin{align*}
     \frac{1}{4} \tilde{\xi}^\top AM^{-1}A^{T}\tilde{\xi}>p^\ast(\tilde{\xi}).
\end{align*}
Now we apply Lemma \ref{lemm:ineq} to go from the Fenchel dual domain to the original domain. For all $x \in \mathbb{R}^{n}$,
\begin{align*}
     x^\top A^{-\top} MA^{-1}x<p(x).
\end{align*}
This implies that the following inequality should hold for every $x \in \mathbb{R}^n$
\begin{align} \label{ineq: q}
    q(x)>x^\top \Bigl(A^{-\top} MA^{-1}-M\Bigr)x
\end{align}
Hence if we choose $q(\cdot)$ s.t it is lower bounded by a quadratic, i.e, $ q(x)>x^\top Zx \quad \forall x$, for some $Z \in \mathbb{R}^{n \times n}$, the condition on positivity of $r(\cdot)$ reduces to $Z>A^{-\top} MA^{-1}-M$. 

Moving one, for the convexity of $r^\ast(\cdot)$, we first observe the following identity as a result of the inverse function theorem and the properties of the Fenchel Dual:
\begin{align*}
    (\nabla^2p^\star(\xi))^{-1}=\nabla^2p[\nabla p^\star(\xi)]
\end{align*}
By the assumption on the convexity $p(\cdot)$ and $M \succ 0$, and the fact that $q(\cdot)$ is convex, we obtain the following identity on the Hessians
\begin{align*}
    M^{-1}\succ2A^{-1}\nabla^2p^\star(2A^{-\top} Mx)A^{-\top} , \quad \forall x
\end{align*}
This conditions is the result of imposing $\nabla^2 r^\ast(\cdot) \succ 0$. This is equivalent to
\begin{align}\label{ineq:M}
    M\prec \frac{1}{2}A^{T}\nabla^2p[\nabla p^\star(2A^{-\top} Mx)]A, \quad \forall x
\end{align}
Taking $\tilde{x}:=\nabla p^\star(2A^{-\top} Mx)$, we observe that since \eqref{ineq:M} holds for every $x$, then
\begin{align*}
     M\prec\frac{1}{2}A^{T}\nabla^2p(\tilde{x})A, \quad \forall (\tilde{x})
\end{align*}
Using the construction of $p(\cdot)$ we arrive at the following necessary and sufficient condition:
\begin{align}\label{cond:conv_r}
     M-\frac{1}{2}A^{T}MA\prec\frac{1}{2}A^{T}\nabla^2q(\cdot)A \quad \forall (\cdot)
\end{align}
We note that the condition in \eqref{cond:conv_r} is equivalent to the condition in \eqref{ineq: q}. This concludes the proof of Theorem \ref{thm:chooseQ}
\end{document}